\documentclass[twocolumn,showpacs,preprintnumbers]{revtex4}
\usepackage{amsmath}
\usepackage{amssymb}
\usepackage{graphicx}
\usepackage{dcolumn}
\usepackage{bm}
\usepackage{color}

\begin{document}

\preprint{APS/123-QED}
\title{Scalable quantum information processing with atomic ensembles and
flying photons}
\author{Feng Mei,$^{1}$ Mang Feng,$^{2,3}${\color{blue}\footnote{
E-mail: mangfeng@wipm.ac.cn}} Ya-Fei
Yu,$^{1,}${\color{blue}\footnote{E-mail: yfyuks@hotmail.com}} and
Zhi-Ming Zhang$^{1,3}${\color{blue}\footnote{E-mail:
zmzhang@scnu.edu.cn}}} \affiliation{$^{1}$Laboratory of Photonic
Information Technology, SIPSE $\&$ LQIT, South
China Normal University, Guangzhou 510006, China\\
$^{2}$State Key Laboratory of Magnetic Resonance and Atomic and
Molecular Physics, Wuhan Institute of Physics and Mathematics,
Chinese Academy of Sciences, Wuhan 430071, China\\
$^{3}$Centre for Quantum Technologies and Department of Physics,
National University of Singapore, 3 Science Drive 2, Singapore
117543, Singapore}
\date{\today }
\date{\today }

\begin{abstract}
We present a scheme for scalable quantum information processing
(QIP) with atomic ensembles and flying photons. Using the Rydberg
blockade, we encode the qubits in the collective atomic states,
which could be manipulated fast and easily due to the enhanced
interaction, in comparison to the single-atom case. We demonstrate
that our proposed gating could be applied to generation of
two-dimensional cluster states for measurement-based quantum
computation. Moreover, the atomic ensembles also function as quantum
repeaters useful for long distance quantum state transfer. We show
the possibility of our scheme to work in bad cavity or in weak
coupling regime, which could much relax the experimental
requirement. The efficient coherent operations on the ensemble
qubits enable our scheme to be switchable between quantum
computation and quantum communication using atomic ensembles.
\end{abstract}

\maketitle

\section{INTRODUCTION}

Recently, much effort has been paid on ensembles of trapped atoms as
promising candidates for quantum state engineering and quantum information
processing (QIP), such as realization of quantum repeater {\color{blue}\cite%
{repeater1,DLCZ,repeater2}}, storage and generation of photonic states {%
\color{blue}\cite{s1,s2,s3,s5,s6,s7}}, and entanglement generation of the
collective degrees of freedom in separate atomic ensembles {\color{blue}\cite%
{entan1,entan2,entan3}}. Due to no need of addressing atoms individually and
to the enhanced interaction with light, the atomic ensemble qubits seem
superior to the single-particle qubits in QIP. However, the interface
between the atomic ensemble and the light based on Raman scattering in above
schemes requires the state of the atomic ensemble to be transformed into a
photon travelling in a well-defined direction within a well-controlled
period of time. More importantly, it is experimentally more challenging to
achieve universal operations for QIP than quantum communication {\color{blue}%
\cite{interface1,interface2,interface3,interface4}} between different atomic
ensemble qubits by interfacing collective atomic excitations with single
photon pulses.

An attractive technique for a nontrivial two-qubit gating between atomic
ensembles has been proposed in {\color{blue}\cite{ensemblegate}}, in which
the dipole-dipole interaction between highly excited Rydberg states blocks
transitions of more than one Rydberg excitation. This is called the Rydberg
blockade {\color{blue}\cite{blockade}}, which has been observed in clouds of
cold atoms {\color{blue}\cite%
{exper1,exper2,exper3,exper4,exper5,exper6,exper7}} as well as in a
Bose-Einstein Condensate {\color{blue}\cite{exper8}}. There have been a
number of proposals to use the Rydberg blockade for various QIP tasks {%
\color{blue}\cite{QIP1,QIP2,QIP3,QIP4,QIP5,QIP6,QIP7}}, most of which,
however, are hard for scaling to a large numbers of qubits required for a
working QIP. One of the main problems for scalability is that the Rydberg
blockade only produces effective interaction within a certain interaction
range. Although it is possible to interconnect two distant qubits by
repeated using swap operations between neighboring qubits, the error
threshold for swap operations would much suppress the scalability {%
\color{blue}\cite{scale}}.

Together with the state-of-the-art techniques in cavity quantum
electrodynamics {\color{blue}\cite{coupling,BEC}} and the recent impressive
experimental advance for Rydberg blockade {\color{blue}\cite{look1,look2}},
we put forward a scalable ensemble-based QIP scheme with the collective
states of the atomic ensembles encoding the qubits and the flying photons as
ancillas. Making use of the single-photon input-output process {\color{blue}%
\cite{Duan1,Duan2}}, we employ optical cavities with each confining
an atomic ensemble under the far-off-resonant interaction, which
could produce a phase flip for each input single-photon pulse, and
thereby could be extended to controlled quantum gating between
different atomic ensemble qubits. We will show that the gating is
insensitive to the Rydberg blockade error and also to the variation
of the coupling rate $g$ even if the atoms are not well localized
within the Lamb-Dicke regime. Our scheme is also robust to the
errors due to the photon loss from the atomic spontaneous emission,
the photon collection and the detection inefficiency, because the
photon loss only reduces the success rate of the gating but has no
affect on the fidelity of the gating under our measurement. In
addition, the building block of our scheme could be readily used to
generate cluster states with two-dimensional (2D) lattice geometry
and to carry out conditional gating between two remote qubits.
Moreover, we will show the availability of our scheme in different
conditions, e.g., in the good or bad cavity and in weak or strong
coupling regime.

Our scheme has following advantages. $\left( i\right) $ The
entanglement between two atomic ensemble qubits is achieved by a
single photon flying sequentially through two cavities confining the
two atomic ensemble qubits respectively. This process is
intrinsically of higher success rate than those based on coincident
detection of two emitting photons going through a polarizing beam
splitter (PBS). $\left( ii\right) $ Compared to the single atom
cases, the enhancement of the coherent coupling by $\sqrt{N}$ in our
scheme, with $N$ the number of the atoms in the ensemble, could
somewhat relax the experimental requirement. This also enables fast
qubit rotation and the efficient readout in our scheme, useful for
measurement-based quantum computation. Our atomic ensembles in the
cavities can also function as good quantum repeaters for
long-distance quantum communication \cite{repeater1,DLCZ,repeater2}.
As a result, quantum computation and quantum communication are
readily switchable with each other in our system. $\left(
iii\right)$ Compared with a previous proposal \cite{QIP5} for
generation of atomic ensemble cluster state, our scheme is more
efficient and the repetition attempts only scale up polynomially
with the atomic ensemble qubit number. $\left( iv\right) $ Our
scheme can work well under wide range of experimental parameters,
where the quantum gating associated with the case of the weak
coupling or large cavity decay is made by means of the Faraday
rotation.

The paper is structured as follows. The next section focuses on the
interaction between a single-photon pulse and an atomic ensemble in
an optical cavity, which yields a controlled phase flip (CPF)
gating. The single-photon pulse going through different spatially
separate cavities could lead to entanglement and quantum gating
between the confined atomic ensembles, as discussed in Section
{\color{blue}III}. We will consider in Sections IV and V the
experimental feasibility of our scheme and the possibility of our
scheme working in the weak coupling regime or in bad cavity. The
last section is for a short summary.

\section{Our IDEA and operations}

\subsection{Rydberg blockade regime and logical qubit encoding}

The Rydberg blockade {\color{blue}\cite{ensemblegate,blockade}}
relies on the interaction between the atoms in the ensemble, which
is intrinsically of the weak $R^{-5}$ or $R^{-6}$ van der Waals type
in the absence of external electric field with $R$ the distance
between the atoms. Under an external electric field, however, the
interaction would be much enhanced because the Rydberg states own
permanent dipole moments $\mu \sim n^{2}ea_{0} $, with $n$ the
principal quantum number, $e$ the electronic charge and $a_{0}$ the
Bohr radius. Unfortunately, if the two atoms $i$ and $j$ are fixed,
it has been found {\color{blue}\cite{blockshift}} that the
interaction would be vanishing when the angle $\theta _{ij\text{ }}$
between the interatomic separation $\mathbf{R}$ and the electric
field $\mathbf{E}$ approaches $\arccos \left( 1/\sqrt{3}\right) $.
In the case of an atomic ensemble with $N $ identical atoms within
the blockade range, however, it is not practical to fully avoid
$\theta _{ij\text{ }}\left( i,j\in \left[ 1,N\right] ,\text{
}j>i\right) =\arccos \left( 1/\sqrt{3}\right) $ for each atoms to
get available Rydberg blockade. Nevertheless, we might employ
F\"{o}ster process, which could yield an isotropic Rydberg atom
interaction of comparable strength $R^{-3}$
{\color{blue}\cite{blockshift}} even in the absence of external
electric field. In this process, if there exists a degeneracy in the
energy level structure, i.e. $nlj+nlj\rightarrow n^{\prime
}l^{\prime }j^{\prime }+n^{\prime \prime }l^{\prime \prime
}j^{\prime \prime }$, the interaction between the Rydberg atoms will
be resonantly enhanced. Despite this, we should avoid choosing the
F\"{o}ster zero states by taking $l^{\prime }=l^{\prime \prime
}=l+1$ and $j^{\prime}=j^{\prime \prime }=j+1$
{\color{blue}\cite{shift}}.

\begin{figure}[h]
\includegraphics[width=8cm]{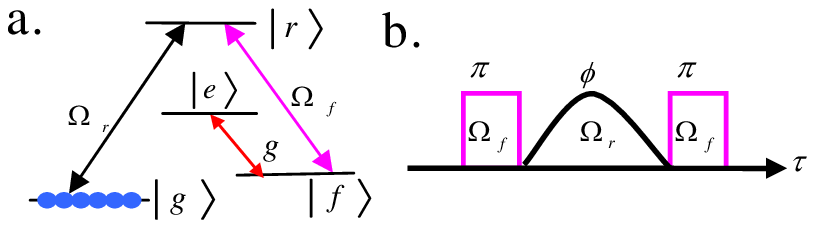}
\caption{(Color online) (a) The relevant level structure of the atoms in the
ensemble. The atomic transition $\left\vert g\right\rangle $ $\left(
\left\vert f\right\rangle \right) \rightarrow \left\vert r\right\rangle $ is
driven by the classical laser with the Rabi frequency $\Omega _{r}\left(
t\right) $ $\left( \Omega _{f}\left( t\right) \right) $, and $\left\vert
f\right\rangle \rightarrow \left\vert e\right\rangle $ is resonantly coupled
to the cavity mode with a coupling rate $g$. (b) A sequence of laser pulses
for a single logical qubit rotation.}
\end{figure}

In our scheme, the Rydberg blockade is utilized to generate the
single excited symmetric atomic state and to rotate the single qubit
state. The interaction for the blockade is at F\"{o}ster resonance.
Fig. {\color{blue} 1(a)} shows the relevant levels of each atom in
the atomic ensemble with the metastable lower states $\left\vert
g\right\rangle $ and $\left\vert f\right\rangle$ for long-time
storage of qubit information and the high-lying Rydberg state
$\left\vert r\right\rangle $ and the excited state $|e\rangle $ for
ancillas. Assuming that all the atoms have been cooled to
micro-Kelvin and prepared in the ground state $\left\vert
g\right\rangle $ in a far off-resonant optical trap (FORT). We
define the logic qubits by the collective atomic states
\begin{eqnarray}
\left\vert 0\right\rangle &=&\otimes _{i=1}^{N}\left\vert g\right\rangle
_{i},  \notag \\
\left\vert 1\right\rangle &=&\left( 1/\sqrt{N}\right)
\sum\nolimits_{i=1}^{N}\left\vert g_{1}...f_{i}...g_{N}\right\rangle \text{.}
\label{1}
\end{eqnarray}
The single qubit rotation in our scheme is implemented by a sequence
of three laser pulses {\color{blue}\cite{QIP1}} as shown in Fig.
{\color{blue} 1(b)}: $\left( i\right) $ The flip operation by a
$2\pi $ pulse, i.e., $\left[ \int d\tau \Omega _{f}\left( \tau
\right) /2=\pi \right] $, results in $\left\vert f\right\rangle
\rightarrow \left\vert r\right\rangle $; $\left( ii\right) $ A
coherent evolution between $\left\vert r\right\rangle $ and
$\left\vert g\right\rangle $ by a $2\phi $ pulse $\left[ \int d\tau
\Omega _{r}\left( \tau \right) /2=\phi \right] $; $\left( iii\right)
$ Another flip operation by a 2$\pi $ pulse results in $\left\vert
r\right\rangle \rightarrow \left\vert f\right\rangle $. In the step
$\left( ii\right) $, the Rydberg blockade guarantees only a single
excitation in the atomic ensemble.

If $\phi =\pi $ or $\pi /2$, the single qubit rotation corresponds to the
Pauli $X$ or Hadamard gate. Moreover, the Rydberg states should be excited
in a Doppler-free fashion, which can be accomplished using two
counter-propagating laser waves. As the logical qubits can be rotated
rapidly {\color{blue}\cite{blockade}} and measured with high efficiency by
resonance fluorescence detection {\color{blue}\cite{QIP2}}, our scheme would
be useful for one-way quantum computation {\color{blue}\cite{measure}}, as
discussed later.

\begin{figure}[tbp]
\includegraphics[width=8cm,height=2.3cm]{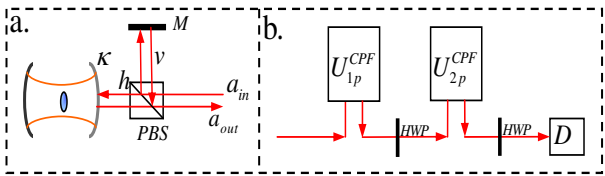}
\caption{(Color online) (a) Schematic setup for the basic building blocks.
The single-photon pulse is injected into the cavity after passing through a
PBS. The optical paths from the PBS to the cavity and to the mirror \textbf{M%
} are assumed to be equal, then the $h$ polarization component reflected
from the cavity can be superposed coherently with the $v$ polarized
component reflected via the mirror. (b) The CPF gate between two qubits. The
box D includes a PBS and two single-photon detectors for the output.}
\end{figure}

\subsection{CPF gating between a single-photon pulse and the atomic ensemble}

Combined with the Rydberg blockade, our basic building block for QIP works
based on the cavity input-output process. As shown in Fig. {\color{blue}2(a)}%
, the atomic ensemble is trapped in a one-sided cavity. The atomic
transition between $\left\vert f\right\rangle $ and $\left\vert
e\right\rangle $ is coupled resonantly to the cavity mode $a_{h}$ and also
resonantly driven by the $h$ polarization component of the input
single-photon pulse. In the rotating frame with respect to the cavity
frequency, the interaction of the atoms with the cavity mode is described by
the Hamiltonian,
\begin{equation}
H=\sum\nolimits_{i=1}^{N} (g_{i}a\sigma _{+}^{i}+g_{i}^{\ast
}a^{\dag }\sigma _{-}^{i}),
\end{equation}%
where $\sigma _{+}^{i}=\left\vert e\right\rangle \left\langle f\right\vert ,$
$\sigma _{-}^{i}=\left\vert f\right\rangle \left\langle e\right\vert ,$ $%
g_{i}$ is the coupling rate between the $i$th atom and the cavity mode. For
simplicity of treatment, we may assume $g_{i}=g$ from now on. By
adiabatically eliminating the cavity mode {\color{blue}\cite%
{inputoutput,Duan1}}, the cavity output $a_{h}^{out}(t)$ corresponds to

\begin{equation}
a_{h}^{out}(t)=\frac{i\Delta -\kappa /2}{i\Delta +\kappa /2}a_{h}^{in}(t)
\end{equation}%
where $\kappa $ is the cavity decay rate, $\Delta $ is the detuning
between the input photon and the dressed cavity mode, and
$a_{h}^{in}(t)$ is the one dimensional input field operator
satisfying $\left[ a_{h}^{in}(t),\text{ } a_{h}^{in+}(t^{\prime
})\right] =\delta (t-t^{\prime })$. If the atomic ensemble is
initially in the state $\left\vert 0\right\rangle $, the
Hamiltonian\ $H$ does not work and thereby $\Delta =0$. So we have
\begin{equation}
a_{h}^{out}(t)=-a_{h}^{in}(t).
\end{equation}
In contrast, if the atomic ensemble is in the state $\left\vert
1\right\rangle $ and the input field senses the dressed cavity modes
$\Delta =\pm g$ {\color{blue}\cite{Duan1}}, in the case that $g\gg
\kappa $, we get $a_{h}^{out}(t)=a_{h}^{in}(t)$. As a result, the
output of the photon means an implementation of the CPF gate
$U_{ap}^{CPF}=e^{i\pi \left\vert 0h\right\rangle \left\langle
0h\right\vert }$ between the single-photon and the atomic ensemble.

The same gating operation could also be achieved in weak coupling
regime or in a bad cavity if we could set the values of the
parameters appropriately. We will discuss this point later.

\begin{figure}[tbp]
\includegraphics[width=4.5cm]{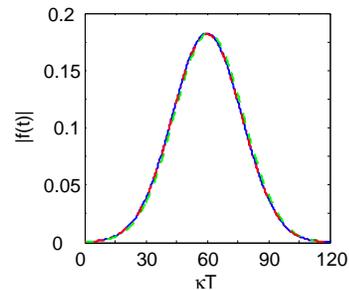}
\caption{(Color online) Pulse shape for the input pulse (solid line)
and the output pulse of the single photon with the atomic ensemble
qubit in the state $\left\vert 0\right\rangle $ (dashed line) and
$\left\vert 1\right\rangle $ (dash-dotted line). The input pulse
shape is assumed to be Gaussian with $f_{in}(t)\propto \exp
[-(t-T/2)^{2}/(T/5)^{2}]$, where $T/5$ is the pulse width. As they
are nearly identical and nearly completely overlapping, the pulses
are hard to be distinguished in the plot.}
\end{figure}

\subsection{The gate performance}

In what follows, to quantitatively characterize the CPF gating, we
will perform a numerical simulation by the method specified in
{\color{blue}\cite {Duan1}} using the realistic parameters $\left(
g_{0}\text{, }\kappa \text{, }\gamma _{s}\right) /2\pi =\left(
34,4.1,2.6\right) $ MHz {\color{blue}\cite {para}}. Assuming the
initial state of the total system to be $\left\vert \Psi
_{in}\right\rangle =\sum\nolimits_{i}c_{ip}\left\vert i\right\rangle
_{a}\left\vert p\right\rangle _{i}$ with $i\in \{0,1\}$ and $p\in
\{h,v\}$, $ \sum\nolimits_{i}\left\vert c_{ip}\right\vert ^{2}=1$,
where $\left\vert p\right\rangle _{i}=\int dtf_{in}\left( t\right)
a_{p}^{in\dag }\left\vert vac\right\rangle $ is the input photon
state with $f_{in}\left( t\right) $ the input Gaussian pulse, we
have the state of the output pulse to be $\left\vert p'\right\rangle
_{i}=\int dtf_{out}^{ip}\left( t\right) a_{p}^{out\dag }\left\vert
vac\right\rangle $ with $f_{out}^{ip}\left( t\right) $ the output
pulse shape. Normally, from the above model, we can directly obtain
the result $f_{out}^{0h}\left( t\right) =-f_{in}\left( t\right) $
and $f_{out}^{ip}\left( t\right) =f_{in}\left( t\right) $ $\left(
ip\in \{0v,1v,1h\}\right) $. Our simulation shown in Fig.
{\color{blue}3} demonstrates that the output pulse shapes
$\left\vert f_{out}^{0h}\left( t\right) \right\vert $ and
$\left\vert f_{out}^{1h}\left( t\right) \right\vert $ overlap very
well with the input pulse shape $f_{in}\left( t\right) $.
Consequently, the fidelity $F$ of the CPF gate is determined by the
shape matching degree, as plotted in Fig. {\color{blue}4(a)} where
the gating is nearly with the fidelity up to $99.7\%$ in the case
that the pulse duration $T$ is much larger than $1/\kappa $.

Furthermore, the gating fidelity is insensitive to the variation of the
coupling rate. Although $g$ varies by a factor of 2 due to the residual
atomic motion in current experiments {\color{blue}\cite{para}}, the change
of the fidelity in our calculation is only about $10^{-3}$. As a result, the
atoms is not necessarily to be strictly localized within the Lamb-Dicke
regime. In our numerical simulation, an imaginary part $-i\frac{\gamma _{s}}{%
2}\sum\nolimits_{i=1}^{N}\left\vert e\right\rangle _{i}\left\langle
e\right\vert $ is introduced to describe the atomic spontaneous emission
rate $\gamma _{s}$ in the Hamiltonian $H$ in Eq. {\color{blue}(2)}. With the
typical choice $g=3\kappa $, the leakage rate for a CPF gating is $%
P_{e}=P_{s}/4=1.7\%$ (see Fig. {\color{blue}4(b)}), which implies a high
success probability of the gating.

\begin{figure}[h]
\includegraphics[width=8cm,height=4cm]{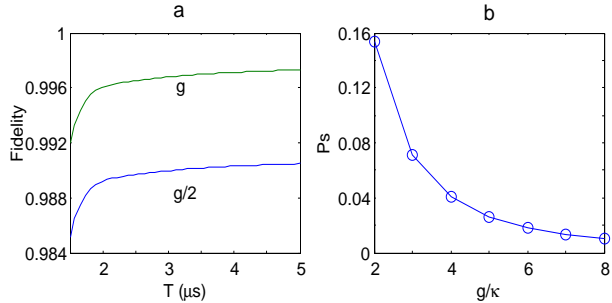}
\caption{(Color online) (a) The gating fidelity versus the pulse duration $T$%
. (b) The photon loss probability $P_{s}$ due to the atomic spontaneous
emission with respect to the coupling rate $g$ in units of $\protect\kappa $%
, where $T=120/\protect\kappa $ is used and $\left( g_{0}\text{, }\protect%
\kappa \text{, }\protect\gamma _{s}\right) /2\protect\pi =\left(
34,4.1,2.6\right) $ MHz.}
\end{figure}

\section{APPLICATION}

\subsection{Conditional gates between atomic ensembles}

The above CPF gating between the atomic ensemble and the
single-photon can be extended to nontrivial two-qubit gating between
atomic ensembles as illustrated in Fig. {\color{blue}2(b)}, where
$U_{ap}^{CPF}$$\left( a=1,2\right) $ box corresponding to the setup
in Fig. {\color{blue}2(a)} functions as the CPF gate between the
atomic ensemble $a$ and the single-photon. The single-photon pulse
injected into the box is initially prepared in the state $\left(
\left\vert h\right\rangle +\left\vert v\right\rangle \right)
/\sqrt{2}$ with $\left\vert h\right\rangle $ and $ \left\vert
v\right\rangle $ the photonic state with polarizations $h$ and $v$,
respectively. The pulse is reflected successively from the two
boxes, with a half-wave plate (HWP) inserted into the optical path
between the two boxes which performs the rotation $\left\vert
h\right\rangle \rightarrow \left(
\left\vert h\right\rangle +\left\vert v\right\rangle \right) /\sqrt{2}$ and $%
\left\vert v\right\rangle \rightarrow \left( \left\vert h\right\rangle
-\left\vert v\right\rangle \right) /\sqrt{2}$. The final output
single-photon pulse after passing through a HWP is detected by its
polarization corresponding to the measurement of the polarization in the
basis $\left\{ \left( \left\vert h\right\rangle \pm \left\vert
v\right\rangle \right) /\sqrt{2}\right\} $.

By a straightforward algebra, one can easily find that, if the
photon is detected in the state $\left\vert h\right\rangle $, the
CPF gating $U_{12}^{CPF}=e^{i\pi \left\vert 11\right\rangle
_{12}\left\langle 11\right\vert }$ succeeds. If the detection is
made on the state $\left\vert v\right\rangle $, the CPF gating could
also succeed after an additional single-qubit operation $\sigma
_{x}$ on the atomic ensemble $2$. As a result, a CNOT gate, with the
CPF gate sandwiched by two Hadamard gates, is available.

\subsection{Cluster state preparation}

Assume the atomic ensembles $1$ and $2$ to be initially prepared in
the state $\left\vert 0\right\rangle _{1}\left\vert 0\right\rangle
_{2}$, and then in the state $\left\vert \varphi _{0}\right\rangle
=$$\left\vert +\right\rangle _{1}\left\vert +\right\rangle _{2}$ by
Hadamard gates with $\left\vert +\right\rangle =\left( \left\vert
0\right\rangle +\left\vert 1\right\rangle \right) /\sqrt{2}$. A CPF
gating between the two atomic ensembles, as demonstrated in Fig.
{\color{blue}2(b)}, yields a cluster state like,
\begin{equation}
\left\vert \varphi \right\rangle _{12}=\frac{1}{2}[(\left\vert
0\right\rangle _{1}+\left\vert 1\right\rangle _{1}\sigma
_{z}^{2})(\left\vert 0\right\rangle _{2}+\left\vert 1\right\rangle _{2})]
\end{equation}
with $\sigma _{z}^{i}=\left\vert 0\right\rangle _{i}\left\langle
0\right\vert -\left\vert 1\right\rangle _{i}\left\langle
1\right\vert $. This is to say, if the output photon is detected as
described in above subsection, we obtain successfully the state in
Eq. ({\color{blue}5}). The idea could be directly extended to
many-qubit case. For instance, two pairs of atomic ensembles $1,2 $
and $3,4$ have been prepared independently in the state $\left\vert
\varphi \right\rangle _{12}\otimes \left\vert \varphi \right\rangle
_{34}$. The two pairs of atomic ensembles could be connected by a
CPF gate between the qubits $2$ and $3$, which yields a four-partite
atomic cluster state,
\begin{equation}
\left\vert \varphi \right\rangle _{1-4}=\frac{1}{4}\otimes
_{i=1}^{4}(\left\vert 0\right\rangle _{i}+\left\vert 1\right\rangle
_{i}\sigma _{z}^{i+1})\text{,}
\end{equation}%
where $\sigma _{z}^{5}\equiv 1$. With the similar idea to above operations,
we can efficiently generate an $n$-party atomic ensemble cluster state
\begin{equation}
\left\vert \varphi \right\rangle _{1-n}=\frac{1}{\sqrt{2^{n}}}\otimes
_{i=1}^{n}(\left\vert 0\right\rangle _{i}+\left\vert 1\right\rangle
_{i}\sigma _{z}^{i+1})\text{,}
\end{equation}%
with the convention $\sigma _{z}^{n+1}\equiv 1$. If the success
probability of the CPF gate is $p$ and the time for each attempt of
the CPF gate is $t_{0}$, the total preparation time for an $n$-qubit
cluster state is $T_{n}\simeq t_{0}(1/p)^{\log _{2}n}$, a polynomial
function of $n$, growing much more slowly than the exponential
scaling in {\color{blue}\cite{QIP5}}. In realistic implementation,
the interface between atomic ensembles and photons with high success
probability favors a high efficiency of cluster state generation.

In fact, the one-dimensional (1D) cluster states are not sufficient
for a universal quantum computation. Recent studies
{\color{blue}\cite {cluster1,cluster2}} have shown a highly
efficient generation of cluster states with any complex 2D lattice
geometry even in the case of low success probability of the CPF
gating. Following the ideas in {\color{blue}\cite
{cluster1,cluster2}}, if we have two sufficiently long cluster
chains which can be done \textit{off-line}, the state in $+$-shape
{\color{blue}\cite {cluster1}} or in star shape
{\color{blue}\cite{cluster2}} with identical legs could be generated
as the building block of the 2D geometry by the shrinking and
removing techniques on the cluster states. Taking the $+$-shape as
an example, via a CPF gate operation, we may fuse the end qubits of
one of the legs of the two $+$-shape cluster states. If it works,
and there are still redundant leg qubits between the two center
qubits in the shapes, we can remove the qubits by applying simple
$X$ measurements on the qubits (i.e., the removing technique).
Simply repeating the procedure, an arbitrary 2D lattice geometry can
be easily constructed. Due to the high success probability of the
CPF gating in our scheme, we can achieve a much more efficient
scaling than in previous schemes {\color{blue}\cite
{cluster1,cluster2}}.

\subsection{Quantum repeater and remotely controlled operation}

Quantum communication over long-distance remains challenging due to
exponential attenuation of the transmitted signals. Fortunately, quantum
repeater could resolve the fiber attenuation problem, reducing the
exponential scaling to polynomial scaling {\color{blue}\cite%
{repeater2,DLCZ,repeater1}}. Following the Duan-Lukin-Cirac-Zoller
model {\color{blue}\cite{DLCZ}}, the quantum repeater can also be
realized in the present system with pairs of atomic ensembles
trapped in separate cavities. By adiabatically changing the Rabi
frequencies of the pumping and repumping pulses with a large
detuning {\color{blue}\cite{DLCZ,adibatic}}, we can store and
retrieve the information at will. Note that the scattered photon
will go to some other optical modes other than the signal mode.
However, when the atomic number $N$ is large, the independent
spontaneous emissions distribute over all the atomic modes, whereas
the contribution to the signal light mode would be small, leading to
a high signal-to-noise ratio $R\sim 4Ng^{2}/\kappa \gamma _{s}$
{\color{blue}\cite{DLCZ}}. As a result, the use of atomic ensembles
as a quantum node could result in collective enhancement. Moreover,
with the high-fidelity entanglement generated by the quantum
repeater, the high-fidelity controlled operation can be realized
between two remote quantum nodes, as shown in Fig. 5.

Besides quantum communication {\color{blue}\cite{DLCZ}}, we show below that
the entanglement between two remote atomic ensembles $E_{1}$ and $E_{2}$ is
useful for quantum gating between two remote atomic ensembles or between two
atoms. This helps for distributed QIP. As an example, we show how to achieve
a CNOT gate between two remote single atoms $S_{1\text{ }}$ and $S_{2}$ (see
Fig. {\color{blue}5}). This can be understood by following identity
\begin{eqnarray}
&&C_{S_{1}E_{1}}C_{E_{2}S_{2}}\left( \left\vert \psi \right\rangle
_{S_{1}S_{2}}\left\vert \varphi \right\rangle _{E_{1}E_{2}}\right)  \notag \\
&=&\sigma _{x}^{S_{2}}C_{S_{1}S_{2}}\left\vert \psi \right\rangle
_{S_{1}S_{2}}\otimes \left\vert 0\right\rangle _{E_{1}}\left\vert
+\right\rangle _{E_{2}}  \notag \\
&&+\left( -\sigma _{z}^{S_{1}}\sigma _{x}^{S_{2}}\right)
C_{S_{1}S_{2}}\left\vert \psi \right\rangle _{S_{1}S_{2}}\otimes \left\vert
0\right\rangle _{E_{1}}\left\vert -\right\rangle _{E_{2}}  \notag \\
&&+C_{S_{1}S_{2}}\left\vert \psi \right\rangle _{S_{1}S_{2}}\otimes
\left\vert 1\right\rangle _{E_{1}}\left\vert +\right\rangle _{E_{2}}  \notag
\\
&&+\sigma _{z}^{S_{1}}C_{S_{1}S_{2}}\left\vert \psi \right\rangle
_{S_{1}S_{2}}\otimes \left\vert 1\right\rangle _{E_{1}}\left\vert
-\right\rangle _{E_{2}},
\end{eqnarray}
where $C_{AB}$ means a CNOT gate on qubit $B$ conditional on $A$,
$\left\vert \varphi \right\rangle _{E_{1}E_{2}}=\left( \left\vert
01\right\rangle _{E_{1}E_{2}}+\left\vert 10\right\rangle
_{E_{1}E_{2}}\right) /\sqrt{2}$, $\left\vert \pm \right\rangle
=\left( \left\vert 0\right\rangle \pm \left\vert 1\right\rangle
\right) /\sqrt{2}$, $\sigma _{z}^{i}$ and $\sigma _{x}^{j}$ denotes
the single qubit Pauli operators acting on the corresponding qubits
$i$ and $j$. It can be easily seen that, the key step for the
nonlocal CNOT gate is the local CNOT gate between atomic ensemble
and single atom qubits, which could be accomplished by the similar
steps in Section III(A). After implementing the local CNOT gate, we
should measure the ensemble qubit $E_{1}$ in the basis $\left\{
\left\vert 0\right\rangle _{E_{1}},\left\vert 1\right\rangle
_{E_{1}}\right\} $ and $E_{2}$ in the basis $\left\{ \left\vert
+\right\rangle _{E_{2}},\left\vert -\right\rangle _{E_{2}}\right\}
$. The measurement results $\left\{ \left\vert 0\right\rangle
_{E_{1}}\left\vert +\right\rangle _{E_{2}},\left\vert 0\right\rangle
_{E_{1}}\left\vert -\right\rangle _{E_{2}},\left\vert 1\right\rangle
_{E_{1}}\left\vert +\right\rangle _{E_{2}},\left\vert 1\right\rangle
_{E_{1}}\left\vert -\right\rangle _{E_{2}}\right\} $ corresponds to
a single qubit operation $\left\{ \sigma _{x}^{S_{2}},I,\sigma
_{z}^{S_{1}}\sigma _{x}^{S_{2}},\sigma _{z}^{S_{1}}\right\} $ on the
single atom qubits. Making use of the atomic ensemble qubits and the
photon-mediated interaction, we could achieve the high-fidelity
remote quantum CNOT gate between two single atoms. As a result, a
high-fidelity long-distance entanglement is mapped to two single
atoms, which can be used for faithful quantum state transfer over
long distance {\color{blue}\cite{QT}}. This nonlocal gating also
provides a basic tool for distributed quantum computation
{\color{blue}\cite{QC}}.
\begin{figure}[tbp]
\includegraphics[width=6cm]{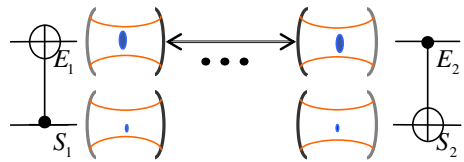}
\caption{(Color online) Schematic setup for CNOT gating between two
remotely located single-atoms.}
\end{figure}

\section{DISCUSSION}

Whether the Rydberg blockade works well or not depends critically on
the weakest interaction between the most apart atoms in the
ensemble. As a result, it is important to choose appropriately the
Rydberg state for carrying out our scheme. At a characteristic
length scale of $R_{c}$, the usual van der Waals interaction could
be treated as the F\"{o}rster interaction. Then the Rydberg--Rydberg
potential energy can be written as $V_{\pm }\left( R\right)
=\frac{\delta }{2}\pm \sqrt{\frac{\delta ^{2}}{4}\pm D_{\varphi
}\frac{C_{3}^{2}}{R^{6}}}$, corresponding to the F\"{o}rster
resonant case $ns_{1/2}+ns_{1/2}\rightarrow np_{3/2}+\left(
n-1\right) p_{2/3}$, the F\"{o}rster defect $\delta =E\left(
np\right) +E\left( \left( n-1\right) p\right) -2E\left( ns\right) $,
the eigenvalue of the Schr\"{o}dinger equation for the van der Waals
interaction with fine structure $ D_{\varphi }=1.333$,
$C_{3}=\frac{e^{2}}{4\pi \epsilon _{0}}\left\langle np\right\vert
\left\vert r\right\vert \left\vert ns\right\rangle \left\langle
\left( n-1\right) p\right\vert \left\vert r\right\vert \left\vert
ns\right\rangle $, $R_{c}=\left( 2C_{3}/\delta \right) ^{1/3}$
\cite{shift}. As shown in Fig. {\color{blue}6(a)}, the $75s$ Rydberg
levels gives $V>100$ MHz of blockade shift at a separation as large
as $5$ $\mu m$. Considering the experimental condition $\Omega
_{r}/2\pi =1$ MHz and $V\gg \Omega _{r}$, we may safely neglect the
triply and higher excited states {\color{blue}\cite{single}}. For
several thousand atoms in the atomic ensemble, the probabilities of
the zero and double excitation are about $10^{-3}\sim 10^{-4}$
{\color{blue}\cite{single}}. We have also calculated the fidelity of
the CPF gate $U_{ap}^{CPF}$ in Fig. {\color{blue} 6(b)} under the
influence of double excitation. We found the fidelity still
remaining as high as $99.2\%$ in the case that the double excitation
probability approaches $P_{2}=0.01$. In contrast, if $P_{2}$ only
varies from $0$ to $0.01$, the fidelity is insensitive to the the
double excitation probability.

\begin{figure}[tbp]
\includegraphics[width=8cm,height=4cm]{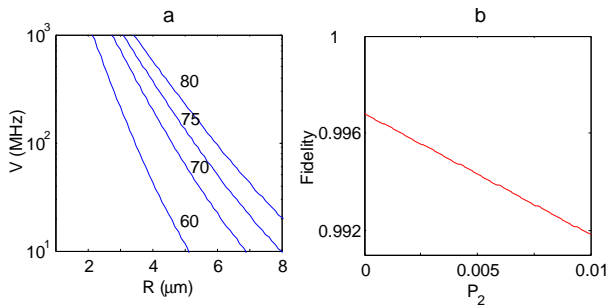}
\caption{(Color online) (a) The dipole-dipole interaction strength $V$ with
respect to atomic separation $R$ for different Rydberg states. (b) The
fidelity of the basic model shown as a function of the double excitation
probability $P_{2}$. The parameters here are employed with the same values
as in Fig. 4.}
\end{figure}

In Section II, we have simply assumed the atoms to be collectively
coupled to the cavity mode with a constant coupling rate $g$.
However, the actual atom-cavity coupling depends on the atom's
position $\mathbf{r}$ through $g\left( \mathbf{r}\right) =g_{0}\cos
\left( k_{c}z\right) \exp \left[ -r_{\perp }^{2}/w_{c}^{2}\right] $,
where $g_{0}$ is the peak coupling rate, $r_{\perp }$ is the radial
distance of the atoms with respect to the cavity axis, $w_{c}$ and
$k_{c}$ are the width and the wave vector of the Gaussian cavity
mode. With current experimental capabilities, the atoms can be
confined inside a potential well along the cavity axis with a nearly
fixed value of $g\left( \mathbf{r}\right) $. But the intracavity
fields of the FORT beam form many potential wells inside the cavity
with different coupling rates in different potential wells. To
implement our scheme to the best, we have to know precisely in which
well the atoms are trapped {\color{blue}\cite{trap}}.

Recent experimental advance {\color{blue}\cite{coupling}} have
achieved many atoms in a cavity with each atom identically and
strongly coupled to the cavity. Based on the fiber-cavity and the
atom-chip technologies, a BEC or cold cloud of $^{87}$Rb atoms in
the $5S_{1/2}\left\vert F=2,m_{F}=2\right\rangle $ ground state can
be prepared and positioned deterministically within the cavity, and
localized entirely within a single antinode of the tight optical
lattice {\color{blue}\cite{coupling}}. For a certain lattice site, a
well-defined and maximal atom-field coupling could be achieved. If
following the definition in {\color{blue}\cite{coupling}} where
$\bar{g}^{2}=\int d\mathbf{r}\rho \left( \mathbf{r}\right)
\left\vert g\left( \mathbf{r}\right) \right\vert ^{2}/N$, with
$g\left( \mathbf{r} \right) $ the position-dependent single-atom
coupling strength, and $\ \rho \left( \mathbf{r}\right) $ the atomic
density distribution, we could calculate the average value of the
single-atom coupling rate $\bar{g}$. For a Gaussian cloud centered
on a single lattice site with $N<1000$ and $\bar{g} /2\pi =200$ MHz,
the fidelity of our CPF gate $U_{ap}^{CPF}$ can approach $ 99.6\%$.
In fact, the combined trap is of the flat disk shape {\color{blue}
\cite{para,disktrap}}, in which the axial trapping frequency $\left(
\nu _{z}\right) $ is much larger than the radial trapping frequency
$\left( \nu _{r_{\perp }}\right) $. By changing the power of the
standing-wave field, we may have $k_{c}\delta z\ll 1$, $\delta
r_{\perp }\ll w_{c}$, then we have negligible variation of the
coupling, i.e., $\delta g/g_{0}\ll 1$. To keep our qubits made of
BEC, however, the axial trapping frequency of the lattice should be
smaller than $\nu _{z}=20$ kHz {\color{blue}\cite{coupling}}.

We have noticed a recent experiment demonstrating the strong
coupling of a $ ^{87}$Rb BEC to a ultrahigh-finesse optical cavity
mode {\color{blue}\cite {BEC}}, in which the atoms in the $^{87}$Rb
BEC occupy a single mode of the matter-wave field and couple
identically to the light field. Inspired by another experimental
advance with the Rydberg excitation of a $^{87}$Rb BEC
{\color{blue}\cite{exper8}}, a $^{87}$Rb BEC may be employed to
encode a qubit with the help of Rydberg blockade. Using the
experimental values $\left( g,k,\gamma _{s}\right) /2\pi =\left(
10.6,1.3,3.0\right) $ MHz {\color{blue}\cite{BEC}}, we have
numerically obtained the fidelity up to $96\%$ in Fig. 4(a) with our
theoretical model. Due to the strong coherence of the BEC, our study
gives rise to fascinating route with the BEC as the qubits for QIP
in the future.

Suppression of the decoherence regarding the collective excitation
is an important issue. In the case that the atom-atom distance is
larger than the reduced optical wavelength of the cavity field
{\color{blue}\cite {ensemblegate}}, i.e. $d=\sqrt{\pi (\delta
r_{\perp })^{2}/N}>\lambda /2\pi $, the collective dephasing rate in
our case is equal to in the single-particle case. Besides, to avoid
the direct interaction between the atoms being in the ground state,
the atom-atom distance should be larger than the radius of the atom
in the ground state, i.e., $r_{g}\sim n_{g}^{2}a_{0}$ with $n_{g}$
the quantum number of the ground state and $a_{0}$ the Bohr radius.
For $n_{g}=5$, $\delta r_{\perp }\sim 5$ $\mu m$, and $N\sim
10^{3}$, we have $\lambda d\sim 1/\pi$ and $d/r_{g}\sim 210$,
implying a valid single-particle approximation. So previous methods
for reducing decoherence in single-atom systems could probably be
used in our case, and the key point for suppressing dephasing is
resorted to a highly stable external magnetic field. In addition,
errors due to the atom-number fluctuations can be ignored if $\delta
N\left( \sim 10\right) \ll N$.

\section{Extension to the WEAK COUPLING AND BAD CAVITY REGIME}

Although the atomic ensemble could in principle expedite the
operations due to the enhanced coupling strength, we have to mention
that the confined atomic ensemble interacting with the flying photon
does not enjoy this advantage because the corresponding coupling
occurs between $|f\rangle$ and the auxiliary level $|e\rangle$. In
this sense, the ensemble qubit interacting with the single flying
photon works as the same as the single atom. As a result, we still
need to work in the high-Q cavity with the strong coupling regime to
accomplish our scheme (See Sec II(B)).

\begin{figure}[h]
\includegraphics[width=5cm]{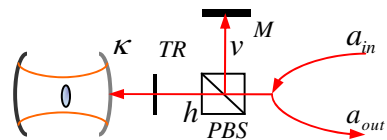}
\caption{(Color online) Schematic setup for implementation of the
atom-photon CPF gate by twice reflections of the single-photon pulse
in the weak coupling regime or in bad cavity. TR is the optical
device exactly controllable for transmitting or reflecting a photon
with very fast switch.}
\end{figure}

Based on a recent publication using Faraday rotation \cite{FENG1},
however, we could extend our scheme to the cavity with low Q factor
or with weak coupling. The key idea is the twice input and output of
the flying single photon with respect to the cavity confining the
atomic ensemble \cite {FENG2}. Since the Faraday rotation is
produced based on the state the atomic ensemble populating, we may
achieve CPF gating under appropriate experimental parameters. For
example, if the atomic ensemble is initially in the state
$\left\vert 1\right\rangle $, once a single-photon pulse is input,
we may have the output single-photon pulse in the state $e^{i\varphi
}\left\vert h\right\rangle $ with $\varphi$ the phase due to the
Faraday rotation. In contrast, if the atomic ensemble is initially
in the state $\left\vert 0\right\rangle $, the single-photon pulse
will sense a far-detuned cavity, yielding $e^{i\varphi
_{0}}\left\vert h\right\rangle $ with $\varphi _{0}$ the Faraday
rotating phase different from $\varphi$ \cite{FENG1}. Supposing
$\omega _{0}=\omega _{c}$, $\omega _{p}=$ $\omega _{c}-\kappa /2$,
with $\omega _{c} $ and $\omega _{p}$ the frequencies of the cavity
and the single-photon, respectively, $\omega _{0}$ the frequency
difference between the levels $\left\vert e\right\rangle $ and
$\left\vert f\right\rangle$ and $\kappa$ the cavity decay rate, if
in the weak coupling or the bad cavity case, i.e. $g=\kappa /2$ and
under the condition of the atoms staying forever in ground states
(also corresponding to the negligible atomic decay rate \cite
{FENG1}), we have the phase shift $\varphi =\pi $ and $\varphi
_{0}=\pi /2$ \cite {FENG1,FENG2}. With the single-photon pulse going
in and out of the cavity twice, as shown in Fig. 7, we may have the
atom-photon CPF gate $U_{ap}^{CPF}=e^{i\pi \left\vert
0h\right\rangle \left\langle 0h\right\vert }$ \cite {FENG2}.

The single photon going through different cavities would yield the
CPF gate between different atomic ensemble qubits \cite{FENG2},
similar to the strong coupling and weak cavity-decay case discussed
in Sec II (B). Therefore, the QIP tasks carried out in good cavities
with strong coupling could also be accomplished in bad cavities or
in weak coupling regime. The enhanced interaction strength due to
the large number of the atoms could improve the efficiency in
accomplishing one-way quantum computing, quantum repeater and
quantum state transfer. More importantly, it makes available to
achieve quantum computation and quantum communication with
sophisticated cavity QED technology \cite{MTR2}.

\section{CONCLUSION}

In conclusion, based on an efficient quantum interface mechanism, we
have shown a scalable ensemble-based QIP scheme. By encoding the
qubits in the atomic ensembles within the Rydberg blockade range, we
could have universal quantum gates with high success probability and
high fidelity. We have also shown that our scheme could work well in
either good or bad cavity and in either strong or weak coupling
regime, which much reduces the experimental requirement.

\section*{ACKNOWLEDGEMENTS}

The authors acknowledge the fruitful discussion with Ming-Sheng
Zhan, Yun-Feng Xiao, Hui Yan, Peng Xu, and Xiao-Dong He. This work
is supported by the National Natural Science Foundation of China
under Grant. Nos. 10404007, 10774163 and 60578055, by the State Key
Development Program for Basic Research of China (Grant No.
2007CB925204 and 2009CB929604), and by NUS Research Grant No.
R-144-000-189-305.

\end{document}